\documentclass[prb,twocolumn,showpacs,floats,eqsecnum,amsmath,amssymb]{revtex4}
\usepackage[dvips]{graphicx}
\def\be{\begin{equation}}
\def\ee{\end{equation}}

\def\bi{\begin{itemize}}
\def\ei{\end{itemize}}
\def\bn{\begin{enumerate}}
\def\en{\end{enumerate}}
\def\bea{\begin{eqnarray}}
\def\eea{\end{eqnarray}}
\def\no{\nonumber}
\def\ba{\begin{array}}
\def\ea{\end{array}}
\def\bd{\begin{displaymath}}
\def\ed{\end{displaymath}}
\def\la{\langle}
\def\ra{\rangle}

\begin{document}



\title{Cumulant expansion for ferrimagnetic spin ($S_1, s_2$) systems}

\author{J. Abouie and A. Langari}

\affiliation{
Institute for Advanced Studies in Basic Sciences,
Zanjan 45195-159, Iran}
\date{\today}

\begin{abstract}

\leftskip 2cm \rightskip 2cm We have generalized the application
of cumulant expansion to ferrimagnetic systems of large spins. We
have derived the effective Hamiltonian in terms of classical
variables for a quantum ferrimagnet of large spins. A
noninteracting gas of ferrimagnetic molecules is studied
systematically by cumulant expansion to second order of ($Js/T$)
where $J$ is the exchange coupling in each molecule, $s$ is the
smaller spin ($S_1, s_2$) and $T$ is temperature. We have observed
fairly good results in the convergent regime of the expansion, i.e
$T > Js$. We then extend our approach to a system of interacting
ferrimagnetic molecules. For one dimensional nearest neighbor
interaction we have observed that the correlation of more than two
neighboring sites is negligible at moderate and high temperature
behavior. Thus the results of a single molecule can be applied to
the chain of interacting molecules for temperatures greater than
classical energy scale, i.e $T>JS_1s_2$. Finally we will discuss
the effect of spin inhomogeneity on the accuracy of this method.

\end{abstract}



\maketitle


\maketitle

\section{Introduction}
Ferrimagnets are systems composed of different spins, mostly of
two types ($S_1, s_2$) \cite{R0, Yamamoto2004}. Many of
ferromagnetic materials are essentially ferrimagnets\cite{R1}. It
is a special class of spin models with much attraction in low
dimension where their universality class is completely different
from homogeneous spin models \cite{Langari}. Antiferromagnetic
chain of spin $s$ models are gapless (gapped) for half-integer
(integer) spins\cite{Haldane}. However ferrimagnetic chains behave
differently. The low energy spectrum of a ferrimagnetic chain
($1,1/2$) is gapless with ferromagnetic order and a gapped band
above it which has antiferromagnetic properties\cite{pati,
mikeska}. This effect causes a crossover from ferromagnetic to
antiferromagnetic behavior at finite temperatures \cite{r4,
yamamoto}.

Spin wave theory gives an explanation for low temperature physics
of ferrimagnetic chains\cite{R3}. However, it is valid for
temperatures smaller than the classical energy scale ($T\lesssim
JS_1s_2$). It is our task to obtain the physics of ferrimagnets
systematically at moderate (where quantum corrections are
important) and higher temperatures. Cumulant expansion (CE)
\cite{fulde, kladko, ma} is our approach to this problem. Recently
this approach has been implemented to study finite temperature
behavior of homogeneous large spin systems\cite{kladko2, Garanin}.
It has been observed that cumulant expansion converges in a region
$T>Js$ which is wider than corresponding one in the high
temperature series expansion and even overlaps the validity regime
of spin wave theory for homogeneous spin systems.

We have generalized the application of cumulant expansion to
ferrimagnetic models. In this approach we obtain a quasiclassical
Hamiltonian for a ($S_1, s_2$) system. The effective Hamiltonian
which is a function of classical variables takes into account the
quantum corrections systematically in the order of ($Js/T$) where $s$
is the smaller value of ($S_1, s_2$). The partition function is
calculated by using the effective Hamiltonian to get physical
properties. Apart from the effective Hamiltonian for a general
($S_1,s_2$) ferrimagnets we found that the n-th order effective
Hamiltonian contains correlation of ($n+2$)-sites.
Moreover, for nearest neighbor  interaction in ferrimagnetic chain, the first order effective
Hamiltonian is composed of 2-sites correlation. Consequently the results of CE for a single
molecule of ($S_1, s_2$) is the same as one dimensional interacting molecules.
We have also found that in the convergence region, the
gas model is a good representation of a chain of
interacting molecules up to 2nd order expansion.

It is remarkable that in the special case $S_1=s_2$, our results match the recent
studies on homogeneous spin chains \cite{kladko2, Garanin}.

The outline of this paper is as follows. In section \ref{sec2} we have introduced
the cumulant expansion in spin coherent state representation.
The effective Hamiltonian and partition function
of two sites gas model are derived in sec. \ref{sec3}. In section \ref{sec4}
we have derived the effective Hamiltonian
of  ferrimagnetic chain and then the physical properties.
Results and discusion have been demonstrainted in sec.\ref{sec5}
where we have presented some physical quantities of a chain with nearest neighbor interaction.


\section{Cumulant expansion in spin coherent state representation\label{sec2}}

The cumulant of N operators or classical variables ($A_i$) is defined as \cite{fulde, kladko}
\bea
&&\la A_{1}\dots A_{n}\ra^c  \no \\
&&=\frac{\partial}
{\partial\lambda_{1}}\dots\frac{\partial}{\partial\lambda_{n}}
\ln\la e^{\lambda_{1}A_{1}+\lambda_{2}A_{2}+\dots}\ra
|_{_{\lambda_{1}=\dots= \lambda_{n}=0}}
\label{ce1}
 \eea
 where $\la\cdots\ra$ means averaging over a classical distribution
function or a quantum state\cite{kladko}. It is easy to show that
in case of all $A_i$ equal to $A$, one will arrive at the
following equality,
\be
 \ln\la e^{\lambda A}\ra=\la e^{\lambda
A}-1 \ra^c.
\label{ce2}
\ee
 To do so, multiplying by
$\lambda^n/n!$ and summing over $n=1,2,...,\infty$ is performed.
Let us show the cumulant of few cases in terms of usual averaging.

\bea
\la A\ra^c&=&\la A\ra,\hspace{5mm}\la A_1 A_2\ra^c=\la
A_1A_2\ra-
\la A_1\ra\la A_2\ra \no \\
\la A_1 A_2 A_3\ra^c&=&\la A_1 A_2 A_3\ra-\la A_1\ra\la A_2 A_3\ra^c
-\la A_2\ra\la A_1 A_3\ra^c \no \\
&-&\la A_3\ra\la A_1 A_2\ra^c-\la A_1\ra\la A_2\ra\la A_3\ra
\label{ce3}
\eea

It is obvious from Eq.(\ref{ce3}) that the cumulant of two or more
operator contains correlations of them. It is a characteristic
property of quantum mechanics that the expectation value of a
product of two operators is different from the product of
expectation values ($\la A_1 A_2\ra-\la A_1\ra\la A_2\ra \neq 0$,
whereas the equality  is valid in classical limit). This is the
correlation of two operators which is inherited from the quantum
state in which the expectation is calculated. So we expect to
obtain a quasiclassical description by using cumulants. In this
respect we are going to obtain an effective (quasiclassical)
Hamiltonian in terms of cumulants which contain quantum
corrections.

The partition function ($Z$) of a spin system can be expressed in
the basis of spin coherent states $|{\bf n}\ra$; i.e. states with the
maximum-spin projection on the axis pointing in the direction of
unit vector ${\bf n}$ [\onlinecite{auerbach}]. The basis of
coherent states are overcomplete, so it contains all quantum
states. The classical state of a spin is achieved in the limit
$S\rightarrow\infty$. The unity operator in this representation is

\be
 {\bf 1}=\frac{2S+1}{4\pi}\int d{\bf n} |{\bf n}\ra\la {\bf n}|.
 \label{ce4}
 \ee
  The trace of an operator in a single spin
problem can be written as:
\bea
tr(A)&=&\sum_m \la
m|A|m\ra=\frac{2S+1}{4\pi}\int d{\bf n}\sum_m
\la m|A|{\bf n}\ra\la {\bf n}|m\ra  \no \\
&=&\frac{2S+1}{4\pi} \int d{\bf n}\la {\bf n}|A|{\bf n}\ra.
\label{ce5}
\eea

So the partition function of a system of $N$-spins defined by the
Hamiltonian $\hat{H}$, is as the following,
\be
Z=\int\Pi_{i=1}^N
\big(\frac{2S_i+1}{4\pi}\big)d{\bf n}_i \la {\bf n}_1 \cdots {\bf
n}_N|e^{-\beta \hat{H}} |{\bf n}_1 \cdots {\bf n}_N\ra
\label{ce6}
\ee
where $\beta=1/T$ (choosing $k_{{\rm B}}=1$). If we define
\be
e^{-\beta {\cal H}}\equiv\la {\bf n}_1 \cdots {\bf n}_N|e^{-\beta
\hat{H}}| {\bf n}_1 \cdots {\bf n}_N\ra
\label{ce7}
\ee
then the
partition function is the same as a classical one. Using
Eq.(\ref{ce2}), the effective Hamiltonian is expressed
in terms of cumulants.
\be
\beta {\cal H}=\la {\bf n}_1 \cdots
{\bf n}_N|1-e^{-\beta \hat{H}}| {\bf n}_1 \cdots {\bf n}_N\ra ^c
\label{ce8}
\ee
The above equation can be expanded in the
following form by a Taylor expansion.
\bea
\beta {\cal H}&=&\beta\la\hat{H}\ra^c-\frac{\beta^2}{2!}\la\hat{H}\hat{H}\ra^c
+\frac{\beta^3}{3!}\la\hat{H}\hat{H}\hat{H}\ra^c+\cdots  \no \\
&=&\beta({\cal H}^{(0)}+{\cal H}^{(1)}+{\cal H}^{(2)}+\cdots)
\label{ce9}
\eea
Thus the effective Hamiltonian (${\cal H}$) can be considered
as a systematic expansion in terms of cumulants of powers of
$\hat{H}$. The first term (${\cal H}^{(0)}$) is the pure
classical contribution and higher orders are responsible for
quantum corrections. In fact this is an expansion in powers of
($Js/T$) where $s$ is the smaller spin of ($S_1, s_2$) in a
ferrimagnetic chain,(it will be shown in next sections). This
expansion is justified whenever $T>Js$.
To the best of our knowledge, a high temperature expansion for
mixed spin systems ($S_1, s_2$) is still missing in the literature. However
to make a qualitative comparison we consider the homogeneous
case \cite{kladko2, Garanin}  ($S_1=s_2=S$). In this case the convergence region of
CE; i.e $T > JS$, is larger than the typical high temperature expansion \cite{blote, rojas}
which is $T > JS^2$.
As we will see in next sections CE can also be
interpreted as an expansion of $1/s$.
So for large spin ferrimagnets the range of convergence is wide.

\section{Two sites gas model \label{sec3}}

Let us first consider an ideal gas-noninteracting-of ferrimagnetic molecules
(gas model) where each molecule composed of two spins ($S_1, s_2$). The
interaction between spins in all molecules is either ferromagnetic
or antiferromagnetic given by the following Hamiltonian.
\be
\hat{H}=\mp J {\bf S}_{1}\cdot {\bf s}_2,~~~~~~~~~~~~J>0
\label{ts1}
\ee
where ${\bf S}_1$ and ${\bf s}_2$ are spin operators of size $S_1$ and
$s_2$ respectively. The ($-$) sign stands for ferromagnetic (F) and
($+$) for antiferromagnetic (AF) interaction. There are two reasons to
study the gas model. Firstly, it is exactly solvable, so we will
compare the results of cumulant expansion with the exact ones to
infer the accuracy of our method. Secondly, we would like to
address the connection of this model to a chain of interacting
ferrimagnetic molecules (next section).


\begin{table}
\caption{The nonzero cumulants of $\langle ({\bf S}_1\cdot {\bf s}_2)({\bf S}_1\cdot {\bf s}_2)\rangle^c$ }
\label{table1}
\begin{ruledtabular}
\begin{tabular}{cccccc}
$X_1$&$x_2$&$X_3$&$x_4$& $\la (X_1x_2)(X_3x_4)\ra^c$& \\
 \hline
 $S_{1}^{+}$&$s_{2}^{+}$&$S_{1}^{-}$&$s_{2}^{-}$&
 $\la S_{1}^{+}S_{1}^{-}\ra^c\la s_{2}^{+}s_{2}^{-}\ra ^c$&$4\omega s^2$\\
 $S_{1}^{+}$&$s_{2}^{z}$&$S_{1}^{-}$&$s_{2}^{z}$&
 $\la S_{1}^{+}S_{1}^{-}\ra^c\la s_{2}^{z}\ra\la s_{2}^{z}\ra $&$2\omega s^3$\\
$S_{1}^{z}$&$s_{2}^{+}$&$S_{1}^{z}$&$s_{2}^{-}$&
 $\la s_{2}^{+}s_{2}^{-}\ra^c\la S_{1}^{z}\ra^c\la S_{1}^{z}\ra$&$2\omega^2s^3$\\
\end{tabular}
\end{ruledtabular}
\end{table}
The eigenstates of Eq.(\ref{ts1}) are labelled by total spin
$S_t=|S_1-s_2|,\dots,S_1+s_2$. Each state is $2S_t+1$ degenerate
whose eigenvalue is
$E_n=\mp\frac{J}{2}[S_t(S_t+1)-S_1(S_1+1)-s_2(s_2+1)]$. In the
incoming calculation of cumulants we define $s\equiv min\{S_1,
s_2\}=s_2$ and the following parameters,
\be
\omega=\frac{S_1}{s_2}>1, \hspace{5mm}\tilde{J}=J s^2. \label{ts2}
\ee
 In order to achieve the cumulants of Eq.(\ref{ce9}) it is
convenient to express the spin operator on each site in the
coordinate system with the $z$ axis along the coherent state
vector $\mathbf{n}_i\equiv\mathbf{n}_i^z$.
\be
\mathbf{S}_i=\sum_{\alpha_i=z,\pm}\mathbf{n}_i^{\alpha_i}S_i^{\alpha_i}
,~~~~~~\mathbf{n}_{\pm}\equiv(\mathbf{n}_x\mp i\mathbf{n}_y)/2
\label{ts3}
\ee


Where $\mathbf{n}_x$ and $\mathbf{n}_y$ are appropriate
transverse basis vectors. Expanding $\la {\bf S}_1\cdot{\bf s}_2\ra$ in terms
of spin components defined in Eq.(\ref{ts3}) the only nonzero term in
coherent state representation is
$\la (S_1^z\mathbf{n}_1)\cdot(s_2^z\mathbf{n}_2)\ra =
(\mathbf{n}_1\cdot\mathbf{n}_2)\la S_1^z s_2^z\ra$.
So the classical contribution to the effective Hamiltonian in Eq.(\ref{ce9})
is: $\la \hat{H}\ra ^c=\mp\tilde{J}\omega(\mathbf{n}_1\cdot\mathbf{n}_2)$.
Quantum corrections to the zeroth order approximation  (classical)
appear in the remaining terms. Keeping the first two corrections one
should calculate the following expressions.
\bea
\la \hat{H}\hat{H}\ra ^c&=&J^2\la({\bf S}_1\cdot{\bf s}_2)({\bf
S}_1\cdot{\bf s}_2)\ra^c
 \\
\la \hat{H}\hat{H}\hat{H}\ra ^c&=&\mp J^3\la({\bf S}_1\cdot{\bf
s}_2)({\bf S}_1\cdot{\bf s}_2)({\bf S}_1\cdot{\bf s}_2)\ra^c
\label{ts4}
\eea
Using Eq.(\ref{ts3}) there are 81 terms in $\la
\hat{H}\hat{H}\ra ^c$ among which three are nonzero. Table
(\ref{table1}) shows the nonzero terms of four operators cumulant.
A similar calculation has been done for $\la
\hat{H}\hat{H}\hat{H}\ra ^c$ to find the nonzero terms. In this
case each cumulant contains the product of 6 operators where
only 15 of them are nonzero.
 Summing up, we will arrive
at the effective Hamiltonian up to second order corrections
($O(1/s^2)$). It is written in terms of classical variables where
quantum effects has been considered.
\bea
\no{\cal H}&=&\tilde{J}\big[\mp\omega(\mathbf{n}_{1}\cdot\mathbf{n}_{2})\\
\no&&-\frac{1}{4}\left(\frac{\beta\tilde{J}}{s}\right)\omega(\omega+1)
\left(1-(\mathbf{n}_{1}\cdot\mathbf{n}_{2})^2\right)\\
\no&&-\frac{1}{8s}\left(\frac{\beta\tilde{J}}{s}\right)\omega(1-\mathbf{n}_{1}
\cdot\mathbf{n}_{2})^2\\
\no&&\mp\frac{1}{12}\left(\frac{\beta\tilde{J}}{s}\right)^2\left(1-(\mathbf{n}_{1}
\cdot\mathbf{n}_{2})^2\right)\times\\
&&\left(\omega^2-(3\omega^2+\omega^3+\omega)(\mathbf{n}_{1}\cdot\mathbf{n}_{2})\right)
\big] \label{ts5}
\eea
The first term shows the classical
contribution which simply represents the energy of coupled
classical spins whose lengths are $S_1$ and $s_2$. Quantum
corrections have a non-Heisenberg form and their structure becomes
more and more complex with increasing order. These corrections are
important in the intermediate temperature where classical
fluctuations are not strong enough to suppress quantum ones. At
very high temperature the classical term is dominant. It is clear
that Eq.(\ref{ts5}) is an expansion in powers of $\beta\tilde{J}/s
=Js/T$, so it is valid as far as $T>Js$.

\begin{figure}
\centerline{\includegraphics[width=9cm,angle=0]{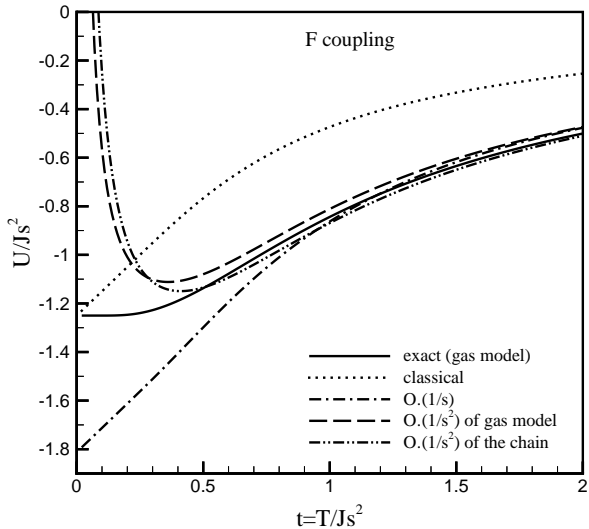}}
\caption{Internal energy per molecule of the ferrimagnetic ($S_1=5/2, s_2=2$) gas model with
ferromagnetic (F) coupling. Exact result of gas model (solid line) and different orders of
cumulant expansion for gas model and chain of interacting molecules.}
\label{fig1}
\end{figure}

\begin{figure}
\centerline{\includegraphics[width=9cm,angle=0]{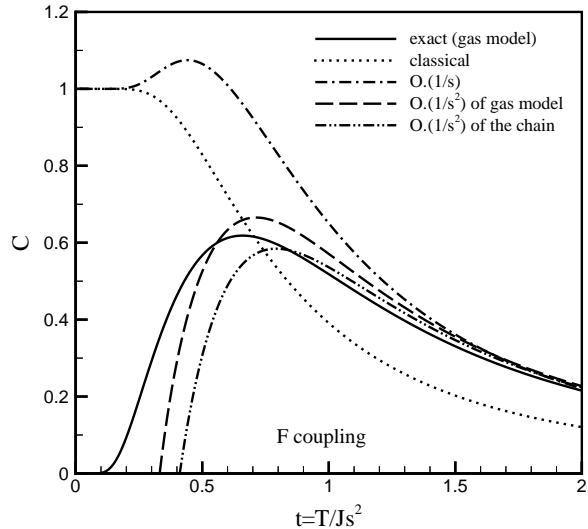}}
\caption{The heat capacity per molecule of the ferrimagnetic ($S_1=5/2, s_2=2$) gas model
with ferromagnetic (F) coupling. Exact result of gas model (solid line) and different orders of
cumulant expansion for gas model and chain of interacting molecules.}
\label{fig2}
\end{figure}

\begin{figure}
\centerline{\includegraphics[width=9cm,angle=0]{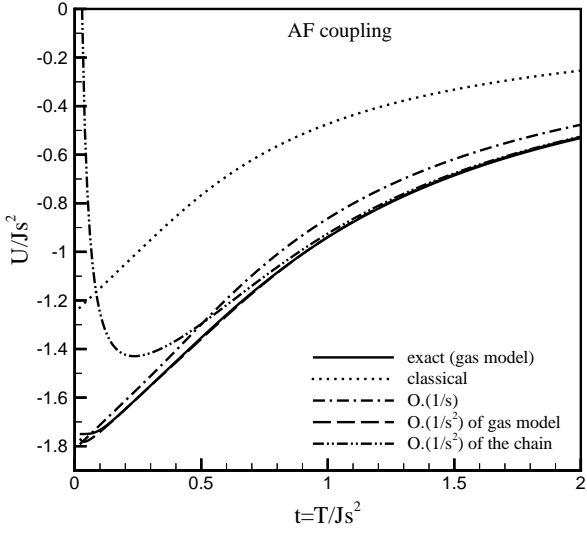}}
\caption{Internal energy per molecule of the ferrimagnetic ($S_1=5/2, s_2=2$) gas model with
antiferromagnetic (AF) coupling. Exact result of gas model (solid line) and different orders of
cumulant expansion for gas model and chain of interacting molecules.}
\label{fig3}
\end{figure}

\begin{figure}
\centerline{\includegraphics[width=9cm,angle=0]{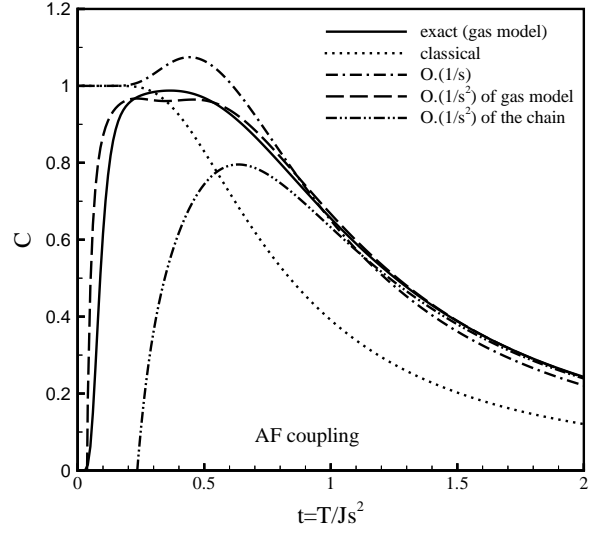}}
\caption{The heat capacity per molecule of the ferrimagnetic ($S_1=5/2, s_2=2$) gas model
with antiferromagnetic (AF) coupling. Exact result of gas model (solid line) and different orders of
cumulant expansion for gas model and chain of interacting molecules.}
\label{fig4}
\end{figure}

To calculate the partition function we use
Eqs.(\ref{ce6}),(\ref{ce7}) where ${\cal H}$ comes from
Eq.(\ref{ts5}). Integration over spherical angles of
$\mathbf{n}_{1}$ and $\mathbf{n}_{2}$ leads to the following
expression.
\bea
\no\ln Z&=&\ln[(2\omega s+1)(2s+1)]+\ln\left(\frac{\sinh(\xi\omega)}{\xi\omega}\right)\\
\no&&+\frac{B(1+\omega)\xi}{2s}+\frac{1}{12s^2}\big(\mp B\omega\xi^2-3\omega\xi^2+15B\xi\\
\no&&-2\omega^2\xi^2+6B\omega\xi-2\xi^2 \pm\frac{6B\xi}{\omega}+3(1+\omega)^2\xi^2\\
&&-\frac{9(1+\omega)^2}{\omega}B\xi-\frac{3B^2\xi^2(1+\omega)^2}{2}\big)
\label{ts6}
\eea
where $\xi=\beta\tilde{J}$ and $B=\coth(\omega\xi)-1/(\omega\xi)$ is the Langevin function.
We can obtain the internal energy ($U=-\partial\ln Z/\partial
\beta$) and heat capacity ($C=\partial U/\partial T$) in this
approximation. Quasiclassical expansions for the internal energy and
heat capacity of this model are shown in
Figs.(\ref{fig1},\ref{fig2},\ref{fig3},\ref{fig4}) for ($S_1,
s_2$)=($5/2, 2$). They show how different contribution of
classical, first and second order corrections sum up to the reasonable
result. We have also plotted the exact solution of ferrimagnetic
gas model for comparison.
The corresponding values for ferrimagnetic chain of interacting molecules have been
also plotted which will be discussed in next sections.
In all of these plots the horizontal
axis is scaled to dimensionless parameter $t=T/Js^2=T/\tilde{J}$.
Thus for large spins, it covers even the intermediate and low temperature
regions.
In the present case the convergence range of cumulant expansion reads to
$t>1/s=1/2$.

A good agreement is observed between second order ($O(1/s^2)$)
cumulant expansion and exact result for internal energy of
F coupling which is shown in Fig.(\ref{fig1}). For
$t<1/2$ it shows large deviation from exact result and diverges as
$t\rightarrow 0$. The reason for divergence is as follows. The
internal energy up to second order approximation ($O(1/s^2)$)
would be,

\bea
\no\frac{U}{Js^2}&=&-\big[\omega B+\frac{1+\omega}{2s}(A\xi+B)\\
\no&+&\frac{1}{12s^2}\big[\big(\mp 2B\omega-6\omega+15A-4\omega^2+6A\omega-4\\
\no&+&\frac{6A}{\omega}+6(1+\omega)^2-9\frac{(1+\omega)^2}{\omega}A-3(1+\omega)^2B^2\big)\xi\\
\no&+&\big(\mp A\omega-3(1+\omega)^2BA\big)\xi^2+15B+6B\omega+\frac{6B}{\omega}\\
&-&9\frac{(1+\omega)^2}{\omega}B\big]\big]
\label{ts7}
\eea
where $A=\partial B/\partial\xi$. As $T\rightarrow 0$, $U$ diverges
since $\xi\rightarrow\infty$. The coefficient of linear term in
$\xi$ comes from both ${\cal H}^{(1)}$ and ${\cal H}^{(2)}$ where
they cancel each other as $\xi\rightarrow\infty$. Whereas the
quadratic term $\xi^2$ needs a contribution from ${\cal H}^{(3)}$
to be regular at $\xi\rightarrow\infty$, which is not present in
second order approximation. The heat capacity for F
coupling of ferrimagnetic molecule is shown in Fig.(\ref{fig2}).
The classical and first order expansion are far from the exact
result, however the correction of second order term makes an
agreement for $t>1/2$.

For AF coupling of gas model the
normalized internal energy ($U/\tilde{J}$) is plotted in
Fig.(\ref{fig3}). It is surprising that the second order expansion
fits very well on the exact result even for very small values of
$t$. A similar behavior can be seen in Fig.(\ref{fig4}) for heat
capacity. To understand the different behavior of cumulant
expansion for F and AF coupling, let us come back to the
energy levels. For F coupling the ground state has
$S_t^{(0)}=(S_1+s_2)$ and any excited state ($E_n$) can be
obtained by $S_t^{(n)}=(S_1+s_2)-n$, $n=1,\dots,2s_2$,
($S_1>s_2$). In the spectrum of F case the difference
of energy levels ($\Delta E^{(l)}=E_l-E_{l-1}=JS_t^{(l-1)}$)
decreases by increasing $E_l$. So it is like a quantum system.
Where the converse is true for AF case. Moreover the absolute
value of $\Delta E$ for the lowest states of F coupling is larger than AF case,
i.e. $\Delta E^{(1)}(F)-\Delta E^{(1)}(AF)=J(2s_2-1)$. This means
that cumulant expansion works better for AF case. In this respect
we should note that a cumulant expansion similar to a high temperature
series expansion starts from $T\rightarrow\infty$, where the
probability of all states are equal. Then different order of
expansion are responsible to recover the non-equal probability of
states. This is crucial when the low energy spectrum is not dense
(like F case) where a big difference exists between the occupation probability
of lowest levels. Thus we expect to observe stronger quantum
effects in the ferromagnetic case.

\section{Ferrimagnetic spin chain \label{sec4}}
We will now consider an interacting system of ferrimagnetic
molecules. As it was noted in the introduction such systems can be
synthesized to be considered effectively as one dimensional models
(chains) \cite{R0, Yamamoto2004}, since the interchain coupling is very small compared to
intra-chain ones. In this case the Hamiltonian of a
ferrimagnetic chain of molecules is:
\be
\hat{H}=\mp\sum_{i,j}^{\frac{N}{2},\frac{N}{2}}J_{2i-1,2j}
\mathbf{S}_{2i-1}\cdot\mathbf{s}_{2j}
\label{ch1}
\ee
It is supposed that larger spins ($S$) sit on the odd-numbered sites and
smaller ones ($s$) on the even sites. The exchange
coupling ($J_{2i-1,2j}$) exists between any different spins. Our
approach is general to cover long-range cases, however, later we
will consider nearest neighbor coupling to obtain physical
quantities.

To find an expansion for the physical quantities we should first
calculate the semiclassical effective Hamiltonian defined in
Eq.(\ref{ce9}). Again, we will consider the cumulant expansion up
to second order corrections of $1/s$ which comes from ${\cal
H}^{(2)}$. Let us first calculate the classical term which is the
simplest one.
\bea
\no\la H\ra^c&=&\la
H\ra=\mp\sum_{i,j}J_{2i-1,2j}\la {\bf S}_{2i-1}\cdot{\bf
s}_{2j}\ra\\
&=&\mp \omega\sum_{i,j}\tilde{J}_{2i-1,2j}{\bf n}_{2i-1}\cdot{\bf
n}_{2j}
\label{ch2}
\eea
where as before
$\tilde{J}_{2i-1,2j}=J_{2i-1,2j}s^2$ and Eq.(\ref{ts3}) has been
used. The first correction ($O(1/s)$) comes from the following
term.
\bea
\la HH\ra^c=\sum_{i,j,l,k}J_{2i-1,2j}J_{2l-1,2k}\la({\bf
S}_{2i-1}\cdot{\bf s}_{2j})({\bf S}_{2l-1}\cdot{\bf
s}_{2k})\ra^c\no\\
\label{ch3}
\eea
There are different sequences of $i$,$j$,$l$ and $k$ in $\la({\bf
S}_{2i-1}\cdot{\bf s}_{2j})({\bf S}_{2l-1}\cdot{\bf s}_{2k})\ra$
where indices can be in the same site, neighboring sites or apart
from each other. We have summarized different cases in appendix
which finally leads to the following result for ${\cal H}^{(1)}$.
\bea
\no{\cal
H}^{(1)}&=&\frac{-\beta\omega}{8s^2}\sum_{i,j}\tilde{J}_{2i-1,2j}^{2}
(1-{\bf n}_{2i-1}\cdot{\bf n}_{2j})^2\\
\no&&-\frac{\beta\omega}{4s}\sum_{i,j,l}\tilde{J}_{2i-1,2j}\tilde{J}_{2i-1,2l}\times\\
\no&&\big({\bf n}_{2j}\cdot{\bf n}_{2l}-({\bf n}_{2i-1}\cdot{\bf n}_{2j})
({\bf n}_{2i-1}\cdot{\bf n}_{2l})\big)\\
&&\no-\frac{\beta\omega^2}{4s}\sum_{i,j,l}\tilde{J}_{2i-1,2j}\tilde{J}_{2l-1,2j}\times\\
&&\no\big({\bf n}_{2i-1}\cdot{\bf n}_{2l-1}-({\bf n}_{2i-1}\cdot{\bf n}_{2j})({\bf
n}_{2l-1}\cdot{\bf n}_{2j})\big)\\
\label{ch4}
\eea
As it is obvious from Eq.(\ref{ch4}), the first
correction which contains quantum effects is not of Heisenberg
type. Moreover the second and third term in Eq.(\ref{ch4})
contains coupled interaction of $3$ sites, i.e. $({\bf
n}_{2i-1}\cdot{\bf n}_{2j})({\bf n}_{2i-1}\cdot{\bf n}_{2l})$ and
$({\bf n}_{2i-1}\cdot{\bf n}_{2j})({\bf n}_{2l-1}\cdot{\bf
n}_{2j})$. This is different from two sites interaction of the
original Hamiltonian, Eq.(\ref{ch1}). It is the price of working
with a classical Hamiltonian instead of the original quantum form.
The last two terms of Eq.(\ref{ch4}) give the information of $3$
sites correlation. We will come back to this point later when
comparing the results of gas model with a chain of interacting
ferrimagnetic molecules. The third term of cumulant expansion is
$\la \hat{H}\hat{H}\hat{H}\ra^c$, where the cumulant of $6$
operators should be calculated. Each cumulant is composed of such
terms, $\la({\bf S}_{2i-1}\cdot{\bf s}_{2j})({\bf
S}_{2l-1}\cdot{\bf s}_{2k})({\bf S}_{2m-1}\cdot{\bf
s}_{2n})\ra^c$. We have shown in appendix the different possible
sequences of ($i,j,l,k,m,n$) which give nonzero cumulant. Summing
up all nonzero cumulants, the second order semiclassical
correction is
\bea
 \no&&{\cal
H}^{(2)}=\mp\frac{\beta^2\omega^2}{4s^2}
\sum_{i,j,l,k}\tilde{J}_{2i-1,2j}\tilde{J}_{2i-1,2l}\tilde{J}_{2k-1,2j}\Theta_{1}\\
\no&\mp&\frac{\omega\beta^2}{12s^2}\sum_{i,j,l,k}\tilde{J}_{2i-1,2j}\tilde{J}_{2i-1,2l}\tilde{J}_{2i-1,2k}\Theta_{2}\\
\no&\mp&\frac{\beta^2\omega^3}{12s^2}\sum_{i,j,l,k}\tilde{J}_{2i-1,2j}\tilde{J}_{2k-1,2j}\tilde{J}_{2l-1,2j}\Theta_{3}\\
\no&\mp&\frac{\beta^2\omega}{24s^3}\sum_{i,j,l}
(\tilde{J}_{2i-1,2j}^{2}\tilde{J}_{2i-1,2l}\Theta_{4}+
\omega\tilde{J}_{2i-1,2j}^{2}\tilde{J}_{2l-1,2j}\Theta_{5})\\
&\mp&\frac{\omega}{24s^4}\sum_{i,j}\tilde{J}_{2i-1,2j}^{3}\Theta_{6}
\label{ch5}
 \eea
where $\Theta_i$ ; $i=1,\dots,6$ are defined below.
\bea
\no\Theta_{1}&=&(\mathbf{n}_{2j}\cdot\mathbf{n}_{2k-1})
-(\mathbf{n}_{2i-1}\cdot\mathbf{n}_{2j})
(\mathbf{n}_{2i-1}\cdot\mathbf{n}_{2k-1})\\
\no&&-(\mathbf{n}_{2j}\cdot\mathbf{n}_{2l})
 (\mathbf{n}_{2k-1}\cdot\mathbf{n}_{2l})\\
\no&&+(\mathbf{n}_{2i-1}\cdot\mathbf{n}_{2j})
(\mathbf{n}_{2i-1}\cdot\mathbf{n}_{2l})
(\mathbf{n}_{2k-1}\cdot\mathbf{n}_{2l})\\
\no&&+\frac{1}{3}\big((\mathbf{n}_{2i-1}\cdot\mathbf{n}_{2k-1})
(\mathbf{n}_{2j}\cdot\mathbf{n}_{2l})\\
\no&&-(\mathbf{n}_{2i-1}\cdot\mathbf{n}_{2l})
(\mathbf{n}_{2j}\cdot\mathbf{n}_{2k-1})\big),\\
\no \Theta_{2}&=&-4(\mathbf{n}_{2i-1}^{+}\cdot\mathbf{n}_{2j})
(\mathbf{n}_{2i-1}\cdot\mathbf{n}_{2l})
(\mathbf{n}_{2i-1}^{-}\cdot\mathbf{n}_{2k})\\
\no&=&(\mathbf{n}_{2i-1}\cdot\mathbf{n}_{2j})
(\mathbf{n}_{2i-1}\cdot\mathbf{n}_{2l})
(\mathbf{n}_{2i-1}\cdot\mathbf{n}_{2k})\\
\no&&-\frac{1}{3}\big[(\mathbf{n}_{2i-1}\cdot\mathbf{n}_{2j})
(\mathbf{n}_{2l}\cdot\mathbf{n}_{2k})\\
\no&&+(\mathbf{n}_{2i-1}\cdot\mathbf{n}_{2l})
(\mathbf{n}_{2j}\cdot\mathbf{n}_{2k})\\
\no&&+(\mathbf{n}_{2i-1}\cdot\mathbf{n}_{2k})
(\mathbf{n}_{2j}\cdot\mathbf{n}_{2l})\big],\\
\no \Theta_{3}&=&-4(\mathbf{n}_{2i-1}\cdot\mathbf{n}_{2j}^{+})
(\mathbf{n}_{2k-1}\cdot\mathbf{n}_{2j})
(\mathbf{n}_{2l-1}\cdot\mathbf{n}_{2j}^{-})\\
\no&=&(\mathbf{n}_{2i-1}\cdot\mathbf{n}_{2j})
(\mathbf{n}_{2k-1}\cdot\mathbf{n}_{2j})
(\mathbf{n}_{2l-1}\cdot\mathbf{n}_{2j})\\
\no&&-\frac{1}{3}\big[(\mathbf{n}_{2i-1}\cdot\mathbf{n}_{2j})
(\mathbf{n}_{2k-1}\cdot\mathbf{n}_{2l-1})\\
\no&&+(\mathbf{n}_{2k-1}\cdot\mathbf{n}_{2j})
(\mathbf{n}_{2i-1}\cdot\mathbf{n}_{2l-1})\\
\no&&+(\mathbf{n}_{2l-1}\cdot\mathbf{n}_{2j})
(\mathbf{n}_{2i-1}\cdot\mathbf{n}_{2k-1})\big],\\
\no \Theta_{4}&=&-\{(\mathbf{n}_{2i-1}\cdot\mathbf{n}_{2l})
\left(1-(\mathbf{n}_{2i-1}\cdot\mathbf{n}_{2j})\right)^2\\
\no&&+2[(\mathbf{n}_{2j}\cdot\mathbf{n}_{2l})+
(\mathbf{n}_{2i-1}\cdot\mathbf{n}_{2j})^2
(\mathbf{n}_{2i-1}\cdot\mathbf{n}_{2l})\\
\no&&-(\mathbf{n}_{2i-1}\cdot\mathbf{n}_{2j})
(\mathbf{n}_{2j}\cdot\mathbf{n}_{2l})\\
\no&&-(\mathbf{n}_{2i-1}\cdot\mathbf{n}_{2j})
(\mathbf{n}_{2i-1}\cdot\mathbf{n}_{2l})]\},\\
\no \Theta_{5}&=&-\{(\mathbf{n}_{2l-1}\cdot\mathbf{n}_{2j})
\left(1-(\mathbf{n}_{2i-1}\cdot\mathbf{n}_{2j})\right)^2\\
\no&&+2[(\mathbf{n}_{2i-1}\cdot\mathbf{n}_{2l-1})-
(\mathbf{n}_{2i-1}\cdot\mathbf{n}_{2j})
(\mathbf{n}_{2i-1}\cdot\mathbf{n}_{2l-1})\\
\no&&-(\mathbf{n}_{2i-1}\cdot\mathbf{n}_{2j})
(\mathbf{n}_{2l-1}\cdot\mathbf{n}_{2j})\\
\no&&+(\mathbf{n}_{2l-1}\cdot\mathbf{n}_{2j})
(\mathbf{n}_{2i-1}\cdot\mathbf{n}_{2j})^2]\},\\
\Theta_{6}&=&(\mathbf{n}_{2i-1}\cdot\mathbf{n}_{2j})
\left(1-(\mathbf{n}_{2i-1}\cdot\mathbf{n}_{2j})\right)^2.
\label{ch6}
\eea
Let us draw your attention to two properties of Eq.(\ref{ch5}). If
we factorize $\tilde{J}=Js^2$ from each term they are in
$(\frac{\beta\tilde{J}}{s})^2$ order. However there are extra
prefactors of powers of ($1/s$) for some terms which make them
less important for large $s$. This is a general property which is
also valid for other orders of cumulant expansion. Generally, in
the cumulant expansion the $n$-th order term, ${\cal H}^{(n)}$,
can be considered of $\tilde{J}(\frac{\beta\tilde{J}}{s})^{n}$
order. The second point comes again from non-Heisenberg type of
interaction in $\Theta_i$. There are some terms which contain
coupled interaction of $4$ different sites, like $({\bf
n}_{2i-1}\cdot{\bf n}_{2j})({\bf n}_{2i-1}\cdot{\bf n}_{2l})({\bf
n}_{2k-1}\cdot{\bf n}_{2l})$ in $\Theta_1$. Such terms give the
information of $4$-point correlation function. Thus the $n$-th
order semiclassical Hamiltonian, ${\cal H}^{(n)}$, recovers the
information of $(n+2)$-sites correlation. However we found that
for nearest neighbor interaction ($J_{i,j}=J\delta_{i,j\pm1}$) the
expectation value of ${\cal H}^{(1)}$ contains only nearest
neighbor interaction. Thus, it contains the correlation of two
sites the same as ${\cal H}^{(1)}$ for gas model. This means the
first order approximation of ferrimagnetic gas model (${\cal
H}^{(0)}+{\cal H}^{(1)}$) would give the same approximation of
first order one dimensional interacting molecules. This will be
shown in next section.


\section{Results and discusions\label{sec5}}
In this section we will present some physical quantities of a
ferrimagnetic chain. Our results of previous section are general
for any long-rang interaction of $J_{ij}$. However we consider
nearest neighbor case, $J_{ij}=J\delta _{i,j\pm 1}$, to obtain
internal energy and heat capacity. Because most of synthesized
materials behave as nearest neighbor interacting molecules, moreover
a comparison with other results is in this case.\\
Similar to the case of gas model the implementation of Eqs.(\ref{ce6}) and (\ref{ce7})
where ${\cal H}$ is approximated by ${\cal H}^{(0)}+{\cal H}^{(1)}+{\cal H}^{(2)}$
leads to

\bea
\no
 Z=\tilde{Z}_{0}[1-\beta\la{\cal H}^{(1)}+{\cal H}^{(2)}\ra+
\frac{\beta^2}{2!}\la[{\cal H}^{(1)}]^2\ra+\dots]
\eea
and
\bea
\no\tilde{Z}_{0}=\left(\frac{2\omega s+1}{4\pi}\right)^{\frac{N}{2}}
\left(\frac{2s+1}{4\pi}\right)^{\frac{N}{2}}Z_{0},\\
\no Z_{0}=\int\prod_{i=1}^{N}dn_{i}~e^{-\beta{\cal H}^{(0)}}.
\eea
To 2nd order approximation the partition function is
\bea
\no Z&\cong&(2\omega s+1)^{\frac{N}{2}}(2s+1)^{\frac{N}{2}}\left(\frac{\sinh(\omega\xi)}{\omega\xi}\right)^{N-1}\times\\
\no&&\bigg\{1+N\left(\frac{\xi^2}{4s^2}(\omega\mp B\omega-\frac{B}{\xi})+\frac{(1+\omega)B\xi}{2s}\right)\\
\no&&+N\bigg[\frac{\xi^3}{4s^2}\left(\pm\frac{2B\omega}{3\xi}+\frac{6B}{\xi^2}-\frac{2\omega}{\xi}\right)\\
&&\no+\frac{\xi^3}{6s^2}\left(-\frac{1}{\xi}+\frac{3B}{\omega\xi^2}-\frac{B^2}{\xi}-\frac{\omega^2}{\xi}
+\frac{3B\omega}{\xi^2}-\frac{B^2\omega^2}{\xi}\right)\bigg]\\
\no&&+N\left(\frac{\xi^2}{8s^2}(1+\omega)^2\left(2-\frac{6B}{\xi\omega}+NB^2\right)
-\frac{\omega\xi^2B^2}{4s^2}\right)\bigg\},\\
\label{r1}
\eea
where we have used some integrals of Ref. [\onlinecite{FISHER}].
Taking derivative of $\ln Z$ to $\beta$ gives the internal energy
\bea
\no\frac{U}{\tilde{J}}&\cong&-\bigg\{\omega B+\frac{1+\omega}{2s}(A\xi+B)\\
\no&&+\frac{1}{12s^2}\big[(2AB\xi^2+2B^2\xi)\left(-2-2\omega^2-3\omega\right)\\
\no&&+(A\xi+B)\big(15+6\omega+\frac{6}{\omega}-\frac{9}{\omega}(1+\omega)^2\big)\\
\no&&+2\xi\left(-3\omega-2-2\omega^2+3(1+\omega)^2\right)\\
&&\mp\omega(A\xi^2+2B\xi)\big]\bigg\}.
\label{r2}
\eea
That's simple to calculate the expression of heat capacity, $C=\partial U/\partial T$, which
is not presented here.

 We have  plotted in Figs. (\ref{fig1}, \ref{fig3}) the internal energy per molecule
($2U/N\tilde{J}$) of ($S_1=5/2, s_2=2$) ferrimagnetic chain ($-\cdot\cdot-\cdot\cdot-$ line)
for F and AF coupling respectively.
As far as we know there is no other result for large spin systems to compare with.
However we compare them with the corresponding values of different orders
of gas model. We observe that the zero
(classical) and first order expansion is the same for gas model
and interacting chain. Because the 1st order expansion (${\cal
H}^{(0)}+{\cal H}^{(1)}$) contains information of two sites
correlation. This can be shown easily by the following equation
\bea
\no\la{\cal H}^{(1)}\ra&=&\frac{-\beta \omega\tilde{J}^2}{8s^2}\sum_{i=1}^N
\la(1-{\bf n}_{i}\cdot{\bf n}_{i+1})^2\ra\\
&&\no-\frac{\beta\tilde{J}^3}{4s}\omega(\omega+1)\sum_{i=1}^{N}\la
1-({\bf n}_{i}\cdot{\bf n}_{i+1})^2\ra.\\
\label{r3}
\eea

As far as the correlation of two sites is concerned
the behavior of a single molecule and a chain of molecules with nearest neighbor
interaction is the same.
The inset of different behavior for chain and non-interacting
molecules comes from the correlation of 3 sites. Such terms exist
in ${\cal H}^{(2)}$ and contribute in $1/s^2$ correction.
We see the difference by $1/s^2$ expansion in
Fig.(\ref{fig1}).
This deviation is more clear for low temperature
regime where quantum effects are important. But
for temperatures greater than classical energy scale,
$T>JS_1s_2$, there is a
good agreement between gas model and ferrimagnetic chain. This
shows that the correlation of more than two sites is important for
low temperatures, $t<\omega$. In other words, at moderate and higher temperature an ideal
gas of molecules represents a chain of ferrimagnets very well.

We have also plotted in Figs.(\ref{fig2}) and (\ref{fig4}) the heat capacity of ferrimagnetic chain
for F and AF coupling respectively.
Similar to the behavior of internal energy, the classical and 1st order ($1/s$) expansion
gives the same results of gas model. The 2nd order correction ($1/s^2$) makes the difference
between chain and single molecule. However a single molecule represents well the behavior
of a chain. This justifies long range correlations can be neglected  for moderate and high
temperature regime of ferrimagnetic chain. This is in agreement with previous conclusion that we
can replace the results of the gas model for a chain of interacting ferrimagnetic molecules
where temperature is greater than classical energy scale, $t>\omega$.

Heat capacity of both Figs.(\ref{fig2}, \ref{fig4}) decreases for high  $t$
which is the sign of antiferromagnetic behavior in the upper part
of spectrum of a ferrimagnetic chain. We also see a Schottky-like peak in $C$
which is the result of ferromagnetic to antiferromagnetic
crossover. However this peak is in the
region where cumulant expansion is not necessarily convergent.
Then our plots for $t\lesssim 1$
might not be reliable, although we know from other arguments \cite{r4, yamamoto}
this peak exists.

To see the effect of spin inhomogeneity we have also considered
the case of $S_1=4, s_2=2$ ($\omega =2$) for both gas model and
ferrimagnetic chain which have not plotted here. As mentioned
before the first order correction contains information of two
sites correlation which makes no difference between gas model and
a chain of interacting molecules. We have found that quantum
effects are more pronounced for larger $\omega$ in F coupling both
in the internal energy and  heat capacity. The second order
cumulant expansion of heat capacity for larger $\omega$  reaches
the exact value for temperature higher than the case of
Fig.(\ref{fig2}). Although we expect a better agreement for lower
temperature by increasing spin magnitude, this is not the case if
we increase one of them. Specially for F coupling, increasing
$\omega$ makes larger energy difference for lower levels which
makes the situation worse.

\begin{figure}[ht!]
\centerline{\includegraphics[width=9cm,angle=0]{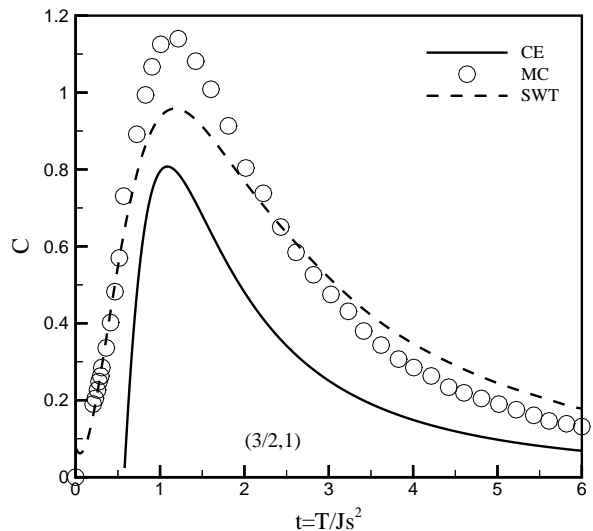}}
\caption{Heat capacity per molecule of ferrimagnetic chain of ($3/2, 1$) with
AF coupling. Second order cumulant expansion (CE), Monte-Carlo (MC) simulation and
spin wave theory (SWT).}
\label{fig5}
\end{figure}


For the AF coupling we have observed almost no different behavior
comparing Figs.(\ref{fig3}, \ref{fig4}) ($\omega =5/4$)  with the
results  of $\omega =2$. In both groups, the second order cumulant
expansion reaches the exact result of gas model at very low
temperature. However this will not remain the same for large
$\omega$. We have increased $\omega$ to $S_1/s_2=7/2$ and observed
good agreement for a little higher temperature. But still fairly
well results of cumulants in the convergent region $t>1/2$.

To the best of our knowledge there is no high temperature expansion for the thermodynamic
properties of ferrimagnets in a closed form in the literature. The large $S$ spin wave approximations
\cite{Yamamoto2004, yamamoto, Takahashi}
are imposed to an extra constraint to be valid for moderate temperature. Because
it is valid for low temperature (by definition) where only the low energy
spectrum has the dominant effect.
Adding an extra constraint needs a self consistent numerical solution to obtain thermodynamic
properties. In comparison, our results are analytic (in a closed form) and general in the sense that
it can be applied to any ferrimagnetic system for arbitrary exchange couplings.
The effective Hamiltonian derived here are in terms of arbitrary $J_{i,j}$ which covers
long range interactions as well as any lattice structures.

However, we finally present the result of second order cumulant expansion (CE),
spin wave theory (SWT) and
Monte-Carlo (MC) simulation \cite{r4}
for heat capacity of ($S_1=3/2, s_2=1$) ferrimagnetic chain in Fig.(\ref{fig5}).
Although cumulant expansion is supposed to work well for large spin systems we present our
results for (3/2, 1) ferrimagnets to compare with available results. The qualitative behavior of
CE is the same as MC simulation but large deviation is observed for moderate temperature.
The main reason for this deviation is the strong quantum nature of small spins. However,
much better agreement is expected for larger spins.

As an out look, this work can be generalized to ladder geometry. It is known that the zero
temperature  behavior of ferrimagnetic ladders \cite{langari2} is different from homogeneous spin
counterparts \cite{dagotto}. It is interesting to see the difference for finite temperature, for instance
the evolution of magnetization plateaux \cite{cabra, langari3}.


\acknowledgments
We would like to thank M. Khorrami and M. R. H. Khajehpour for their
fruitful discusions and comments.


\begin{table}
\caption{Nonzero cumulant of six operators in a chain of interacting molecules (Eq.(\ref{ch1})).
The first column shows the position of different spins. For instance $135,24,6$ shows that
$X_1, X_3$ and $X_5$ sit at a single site, $x_2$ and $x_4$ at the other site and
$x_6$ sits at a separate one. $\la A \ra^c=\la(X_{1}x_{2})(X_{3}x_{4})(X_{5}x_{6})\ra^c$.}
\label{table2}
\begin{ruledtabular}
\begin{tabular}{cccccccc}
&$X_{1}$&$
x_{2}$&$X_{3}$&$x_{4}$&$X_{5}$&$x_{6}$&$\la A \ra^c$\\
\hline
$135,24,6$&$S_{2i-1}^{+}$&$s_{2j}^{+}$&$S_{2i-1}^{z}$&$s_{2j}^{-}$&$S_{2i-1}^{-}$&$s_{2l}^{z}$
&$4\omega^2s^4$\\
&$S_{2i-1}^{z}$&$s_{2j}^{+}$&$S_{2i-1}^{+}$&$s_{2j}^{-}$&$S_{2i-1}^{-}$&$s_{2l}^{z}$
&$4\omega^2s^4$\\
&$S_{2i-1}^{+}$&$s_{2j}^{+}$&$S_{2i-1}^{z}$&$s_{2j}^{-}$&$S_{2i-1}^{-}$&$s_{2l}^{z}
$&$-4\omega s^3$\\
&$S_{2i-1}^{+}$&$s_{2j}^{z}$&$S_{2i-1}^{z}$&$s_{2j}^{z}$&$S_{2i-1}^{-}$&$s_{2l}^{z}$
&$-2\omega s^4$\\
$135,26,4$&$S_{2i-1}^{+}$&$s_{2j}^{+}$&$S_{2i-1}^{-}$&$s_{2l}^{z}$&$S_{2i-1}^{z}$&$s_{2j}^{-}$
&$\omega^2s^4$\\
&$S_{2i-1}^{z}$&$s_{2j}^{+}$&$S_{2i-1}^{+}$&$s_{2l}^{z}$&$S_{2i-1}^{-}$&$s_{2j}^{-}$
&$4\omega^2s^4$\\
&$S_{2i-1}^{+}$&$s_{2j}^{+}$&$S_{2i-1}^{z}$&$s_{2l}^{z}$&$S_{2i-1}^{-}$&$s_{2j}^{-}$
&$-4\omega s^3$\\
&$S_{2i-1}^{+}$&$s_{2j}^{z}$&$S_{2i-1}^{z}$&$s_{2l}^{z}$&$S_{2i-1}^{-}$&$s_{2j}^{z}$
&$-2\omega s^4$\\
$135,46,2$&$S_{2i-1}^{+}$&$s_{2l}^{z}$&$S_{2i-1}^{z}$&$s_{2j}^{+}$&$S_{2i-1}^{-}$&$s_{2j}^{-}$
&$4\omega^2s^4$\\
&$S_{2i-1}^{+}$&$s_{2l}^{z}$&$S_{2i-1}^{-}$&$s_{2j}^{+}$&$S_{2i-1}^{z}$&$s_{2j}^{-}$
&$4\omega^2s^4$\\
&$S_{2i-1}^{+}$&$s_{2l}^{z}$&$S_{2i-1}^{z}$&$s_{2j}^{+}$&$S_{2i-1}^{-}$&$s_{2j}^{-}$
&$-4\omega s^3$\\
&$S_{2i-1}^{+}$&$s_{2l}^{z}$&$S_{2i-1}^{z}$&$s_{2j}^{z}$&$S_{2i-1}^{-}$&$s_{2j}^{z}$
&$-2\omega s^4$\\
$246,13,5$&$S_{2i-1}^{+}$&$s_{2j}^{+}$&$S_{2i-1}^{-}$&$s_{2j}^{z}$&$S_{2k-1}^{z}$&$s_{2j}^{-}$
&$4\omega^2s^4$\\
&$S_{2i-1}^{+}$&$s_{2j}^{z}$&$S_{2i-1}^{-}$&$s_{2j}^{+}$&$S_{2k-1}^{z}$&$s_{2j}^{-}$
&$4\omega^2s^4$\\
&$S_{2i-1}^{+}$&$s_{2j}^{+}$&$S_{2i-1}^{-}$&$s_{2j}^{z}$&$S_{2k-1}^{z}$&$s_{2j}^{-}$
&$-4\omega^2 s^3$\\
&$S_{2i-1}^{z}$&$s_{2j}^{+}$&$S_{2i-1}^{z}$&$s_{2j}^{z}$&$S_{2k-1}^{z}$&$s_{2j}^{-}$
&$-2\omega^3 s^4$\\
$246,15,3$&$S_{2i-1}^{+}$&$s_{2j}^{+}$&$S_{2k-1}^{z}$&$s_{2j}^{-}$&$S_{2i-1}^{-}$&$s_{2j}^{z}$
&$4\omega^2s^4$\\
&$S_{2i-1}^{+}$&$s_{2j}^{z}$&$S_{2k-1}^{z}$&$s_{2j}^{+}$&$S_{2i-1}^{-}$&$s_{2j}^{-}$
&$4\omega^2s^4$\\
&$S_{2i-1}^{+}$&$s_{2j}^{+}$&$S_{2k-1}^{z}$&$s_{2j}^{z}$&$S_{2i-1}^{-}$&$s_{2j}^{-}$
&$-4\omega^2s^3$\\
&$S_{2i-1}^{z}$&$s_{2j}^{+}$&$S_{2k-1}^{z}$&$s_{2j}^{z}$&$S_{2i-1}^{z}$&$s_{2j}^{-}$
&$-2\omega^3 s^4$\\
$246,35,1$&$S_{2k-1}^{z}$&$s_{2j}^{+}$&$S_{2i-1}^{+}$&$s_{2j}^{z}$&$S_{2i-1}^{-}$&$s_{2j}^{-}$
&$4\omega^2s^4$\\
&$S_{2k-1}^{z}$&$s_{2j}^{+}$&$S_{2i-1}^{+}$&$s_{2j}^{-}$&$S_{2i-1}^{-}$&$s_{2j}^{z}$
&$4\omega^2s^4$\\
&$S_{2k-1}^{z}$&$s_{2j}^{+}$&$S_{2i-1}^{+}$&$s_{2j}^{z}$&$S_{2i-1}^{-}$&$s_{2j}^{-}$
&$-4\omega^2s^3$\\
&$S_{2k-1}^{z}$&$s_{2j}^{+}$&$S_{2i-1}^{z}$&$s_{2j}^{z}$&$S_{2i-1}^{z}$&$s_{2j}^{-}$
&$-2\omega^3s^4$
\end{tabular}
\end{ruledtabular}
\end{table}

\appendix

\section{The cumulant of four and six operators for a spin chain}
In order to achieve the first order cumulant i.e. $\la ({\bf
S}_{2i-1}\cdot{\bf s}_{2j})({\bf S}_{2l-1}\cdot{\bf
s}_{2k})\ra^c$, there are three nonzero sequences of $i$, $j$, $l$
and $k$ as follows \bi \item{Nonzero cumulant for $i=l$ and $j=k$.
These are listed in table \ref{table1}}
\item{For $i=l$ and $j\neq
k$, the cumulant is $\la ({\bf S}_{2i-1}\cdot{\bf s}_{2j}) ({\bf
S}_{2l-1}\cdot{\bf s}_{2j})\ra^c$ and its value equals to
$2\omega^2 s^3 ({\bf n}_{2i-1}\cdot {\bf n}_{2j}^{+})({\bf
n}_{2l-1}\cdot {\bf n}_{2j}^{-})$ }
\item{In the case of $j=k$ and
$j\neq l$ the cumulant is $\la ({\bf S}_{2i-1}\cdot{\bf s}_{2j})
({\bf S}_{2i-1}\cdot{\bf s}_{2k})\ra^c$ which is equal to $2\omega
s^3 ({\bf n}_{2i-1}^{+}\cdot {\bf n}_{2j})({\bf n}_{2i-1}^{-}\cdot
{\bf n}_{2l})$} \ei

In order to achieve the second order cumulant of the spin chain
i.e. $\la ({\bf S}_{2i-1}\cdot{\bf s}_{2j})({\bf S}_{2l-1}\cdot{\bf s}_{2k})
({\bf S}_{2m-1}\cdot{\bf s}_{2n})\ra^c$, there are four nonzero sequences
of $i$, $j$, $l$, $k$, $m$ and $n$
as follows
\bn
\item{Three operators sit at a site and the others at another site, the cumulant is
$\la ({\bf S}_{2i-1}\cdot{\bf s}_{2j})({\bf S}_{2i-1}\cdot{\bf s}_{2j})
({\bf S}_{2i-1}\cdot{\bf s}_{2j})\ra^c$. There are only 15 different sequences
which are nonzero.}
\item{Three operators sit at a site, two operators sit at
another one and the sixth one sits at a separate site,
the nonzero cumulants are listed in table \ref{table2}. At the first column of the table \ref{table2},
we can observe operators which can be placed
at the same site, (i.e at the first four row of the table $i=l=k$, $j=l$ and $k\neq n$)
 }
\item{When two operators sit at a site, the another two at another site and the fifth and sixth
one sit at two different sites, the cumulant results to
\bea
\no&&4\omega^2s^4\big(
2({\bf n}_{2i-1}^{+}\cdot{\bf n}_{2j}^{+})({\bf n}_{2i-1}^{-}\cdot{\bf n}_{2l})({\bf n}_{2k-1}\cdot{\bf n}_{2j}^{-})\\
\no&&~~~~~+({\bf n}_{2i-1}^{+}\cdot{\bf n}_{2l})({\bf n}_{2i-1}^{-}\cdot{\bf n}_{2j}^{+})({\bf n}_{2k-1}\cdot{\bf n}_{2j}^{-})\\
\no&&~~~~~+2({\bf n}_{2i-1}^{+}\cdot{\bf n}_{2l})({\bf n}_{2k-1}\cdot{\bf n}_{2j}^{+})({\bf n}_{2i-1}^{-}\cdot{\bf n}_{2j}^{-})\\
\no&&~~~~~+({\bf n}_{2k-1}\cdot{\bf n}_{2j}^{+})({\bf n}_{2i-1}^{+}\cdot{\bf n}_{2j}^{-})({\bf n}_{2i-1}^{-}\cdot{\bf n}_{2l})
\big).
\eea
}
\item{Finally, if three operators sit at a site and other ones sit at three different sites, then
there are two nonzero terms as follows
\bn
\item{If $i=l=m$, $j$, $k$ and $n$ are not equal, the value of cumulant is\\
$-2\omega s^4({\bf n}_{2i-1}^{+}\cdot{\bf n}_{2j})({\bf n}_{2i-1}\cdot{\bf n}_{2l})({\bf n}_{2i-1}^{-}\cdot{\bf n}_{2k})$.}
\item{If $j=k=n$, $i$, $l$ and $m$ are not equal, the value of cumulant is\\
$-2\omega^3 s^4({\bf n}_{2i-1}\cdot{\bf n}_{2j}^{+})({\bf n}_{2k-1}\cdot{\bf n}_{2j})({\bf n}_{2l-1}\cdot{\bf n}_{2j}^{-})$.}
\en
}
\en
Note that, the second order cumulant expansion is the sum of above mentioned cases.



\end{document}